\documentclass[11pt]{iopart}
\pdfoutput=1
\expandafter\let\csname equation*\endcsname\relax
  \expandafter\let\csname endequation*\endcsname\relax

\usepackage{amsmath}
\usepackage{graphicx,eucal}
\usepackage{hyperref}
\usepackage{color}
\usepackage{float}
\usepackage{cite}
\usepackage{subfigure}
\usepackage{url}

\begin{document}

\title{Residence Time Near an Absorbing Set}
  
\author{J. Randon-Furling} \address{SAMM (EA 4543), Universit\'e Paris-1
  Panth\'eon-Sorbonne, 90 rue de Tolbiac,
  75013 Paris, France}

\author{S. Redner} \address{Santa Fe Institute, 1399 Hyde Park Road, Santa
  Fe, New Mexico 87501, USA}
  
\begin{abstract}

  We determine how long a diffusing particle spends in a given spatial range
  before it dies at an absorbing boundary.  In one dimension, for a particle
  that starts at $x_0$ and is absorbed at $x=0$, the average residence time
  of the particle in the range $[x,x+dx]$ is $T(x)=\frac{x}{D}\,dx$ for
  $x<x_0$ and $\frac{x_0}{D}\,dx$ for $x>x_0$, where $D$ is the diffusion
  coefficient.  We extend our approach to biased diffusion, to a particle
  confined to a finite interval, and to general spatial dimensions.  We then
  use the generating function technique to derive parallel results for the
  average number of times that a one-dimensional symmetric nearest-neighbor
  random walk visits site $x$ when the walk starts at $x_0=1$ and is absorbed
  at $x=0$.  We also determine the distribution of times when the random walk
  first revisits $x=1$ before being absorbed.

\end{abstract}

\section{Introduction}

\begin{figure}[ht]
  \centerline{\includegraphics[width=0.65\textwidth]{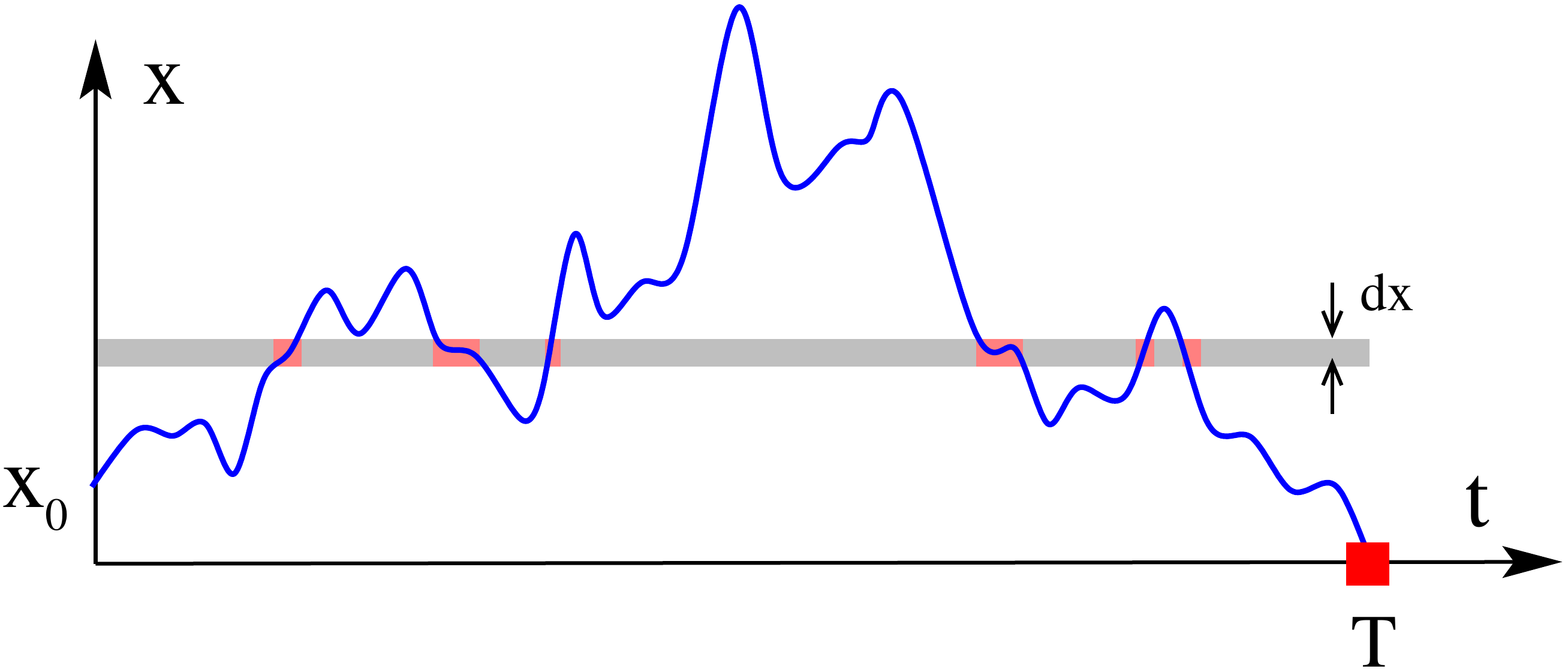}}
  \caption{Schematic trajectory of a diffusing particle in space time that
    starts at $x_0$ and is absorbed at time $T_0$ (square).  The time $T(x)$
    that the particle spends in $[x,x+dx]$ is indicated by the colored
    segments.}
\label{space-time}  
\end{figure}

Suppose that a diffusing particle in one dimension starts at $x_0>0$ and is
absorbed, or equivalently, dies, when $x=0$ is reached.  One classic property
of diffusion is that the particle is sure to eventually reach the origin, but
the average time for this event to occur is
infinite~\cite{F64,S64,MW65,MP10}.  This dichotomy between certain return and
an infinite return time is the source of rich phenomenology and
counter-intuitive phenomena about the statistical properties of diffusion.
Another important feature of diffusion is the shape of its trajectory in
space time (Fig.~\ref{space-time}).  Consider a Brownian particle that starts
at $x_0$ at $t=0$ and first returns to $x=0$ at time $T_0$.  In the
interesting case of $T_0\gg x_0^2/D$, the particle wanders over a large
spatial range before its eventual demise.  A trajectory that stays strictly
in the range $x>0$ until absorption at time $T_0$ is known as a
\emph{Brownian excursion} when the starting point $x_0$ is also equal to
zero~\cite{Chung76}.

Two basic questions about an excursion are: (i) What is its
shape~\cite{BCC03,CBC04}?  (ii) How much time does the excursion spend in the
range $[x,x+dx]$ before being absorbed?  We term this latter quantity as the
residence time.  The properties of the residence time have been addressed in
the mathematics literature by local-time theorems~\cite{L48,MP10,R63,K63}
that specify the time that a Brownian particle spends in the region
$[x,x+dx]$ before being absorbed when the origin is reached.  When $x<x_0$,
the distribution of this residence time was shown to be related to the
distribution of the radial distance of a two-dimensional Brownian
motion~\cite{R63,K63}.  If the particle wanders in a finite domain with
reflection at the domain boundary and absorption at a given point (or points)
within the domain, the residence time at each site is related to the
first-passage time to the absorbing set~\cite{CTB07,BD09,BV14}.  This general
formalism allows one to compute both the average residence time and the
distribution of residence times at a given location.

While the consequences of local-time theorems are profound, the mathematical
literature is sometimes presented in a style that is not readily accessible
to the community of physicists who study random walks, and some of the
results derived in Refs.~\cite{CTB07,BD09,BV14} are extremely general in
their formulation.  In this work, we investigate residence-time phenomena for
both continuum diffusion and the discrete random walk by using simple ideas
and approaches from first-passage processes.  We focus on cases where the
particle is eventually absorbed at a specified boundary (\textit{e.g.}, one
specific side of an interval) and/or starts close to this boundary.  Some of
these situations have been treated previously in Ref.~\cite{A84}, by a more
formal approach than that presented here.

In Sec.~\ref{subsec:si}, we first derive the average residence time within
the interval $[x,x+dx]$ in continuum diffusion, by solving the relevant
diffusion equation.  We extend this approach to: (a) biased diffusion on the
semi-infinite line (Sec.~\ref{subsec:bias}), and (b) unbiased diffusion in a
finite domain $[0,L]$, with the condition that the particle is eventually
absorbed at $x=0$ (Sec.~\ref{subsec:finite}).  We then determine the average
residence time in general spatial dimension in the domain exterior to an
absorbing sphere of radius $a$ (Sec.~\ref{subsec:gend}).

We then turn to the corresponding discrete system of a nearest-neighbor
symmetric random walk that starts at $x_0=1$ and is absorbed when $x=0$ is
reached.  The analog of the residence time is the number of times that the
walk revisits a given point $x$ before it dies at $x=0$.  We write the total
number of steps of this random walk---which is necessarily odd---as $2n+1$.
In Sec.~\ref{sec:rw}, we use generating function methods to derive the
average number of visits to a given site for a symmetric random walk on the
semi-infinite line.  For fixed $n$, we will show that the average number of
times that $x=1$ is revisited equals $3n/(n+2)$.  By averaging this quantity
over the total number of steps of the walk, we will show that there are, on
average, 2 revisits to $x=1$.  Moreover, the average number of times that the
random walk visits a site at $x>1$ equals 4 for any $x> 1$.  These results
match those found in continuum diffusion in the analogous geometry.  Finally,
in Sec.~\ref{sec:first}, we determine the time when a walk first revisits
$x=1$, when it starts at $x=1$ and is eventually absorbed at $x=0$.  We give
some concluding comments in Sec.~\ref{sec:disc}.

\section{Residence Time for Diffusion}
\label{sec:diff}

\subsection{Isotropic diffusion on the semi-infinite line}
\label{subsec:si}

Consider a particle with diffusion coefficient $D$ that starts at $x_0$ and
is absorbed when $x=0$ is reached.  For such a particle, the image method
gives the probability density for the particle to be at position $x>0$
as~\cite{D82,R01}
\begin{align}
  \label{P}
  P(x,t)=\frac{1}{\sqrt{4\pi Dt}}\,\,
  \left[e^{-(x-x_0)^2/4Dt}- e^{-(x+x_0)^2/4Dt}\right]\,.
\end{align}
The time $T(x)$ that the particle, which starts at $x_0$, spends in the range
$[x,x+dx]$ before being absorbed at $x=0$ (Fig.~\ref{space-time}) is simply
the integral of the probability density over all time times $dx$ (see
Refs.~\cite{S64,MW65,FZ01,FZ04,FZ06} for related approaches).  Performing
this integral, with $P(x,t)$ from Eq.~\eqref{P}, the residence time $T(x)$ is
given by
\begin{align}
\label{Nfree}
  T(x)=dx\int_0^\infty \!dt\, P(x,t)=
  \begin{cases}
    {\displaystyle \frac{x}{D}}\,\, dx & \qquad x<x_0\,,\\[5mm]
    {\displaystyle \frac{x_0}{D}}\,\, dx & \qquad x>x_0\,.
  \end{cases}
\end{align}

\begin{figure}[ht]
  \centerline{\subfigure[]{\includegraphics[width=0.425\textwidth]{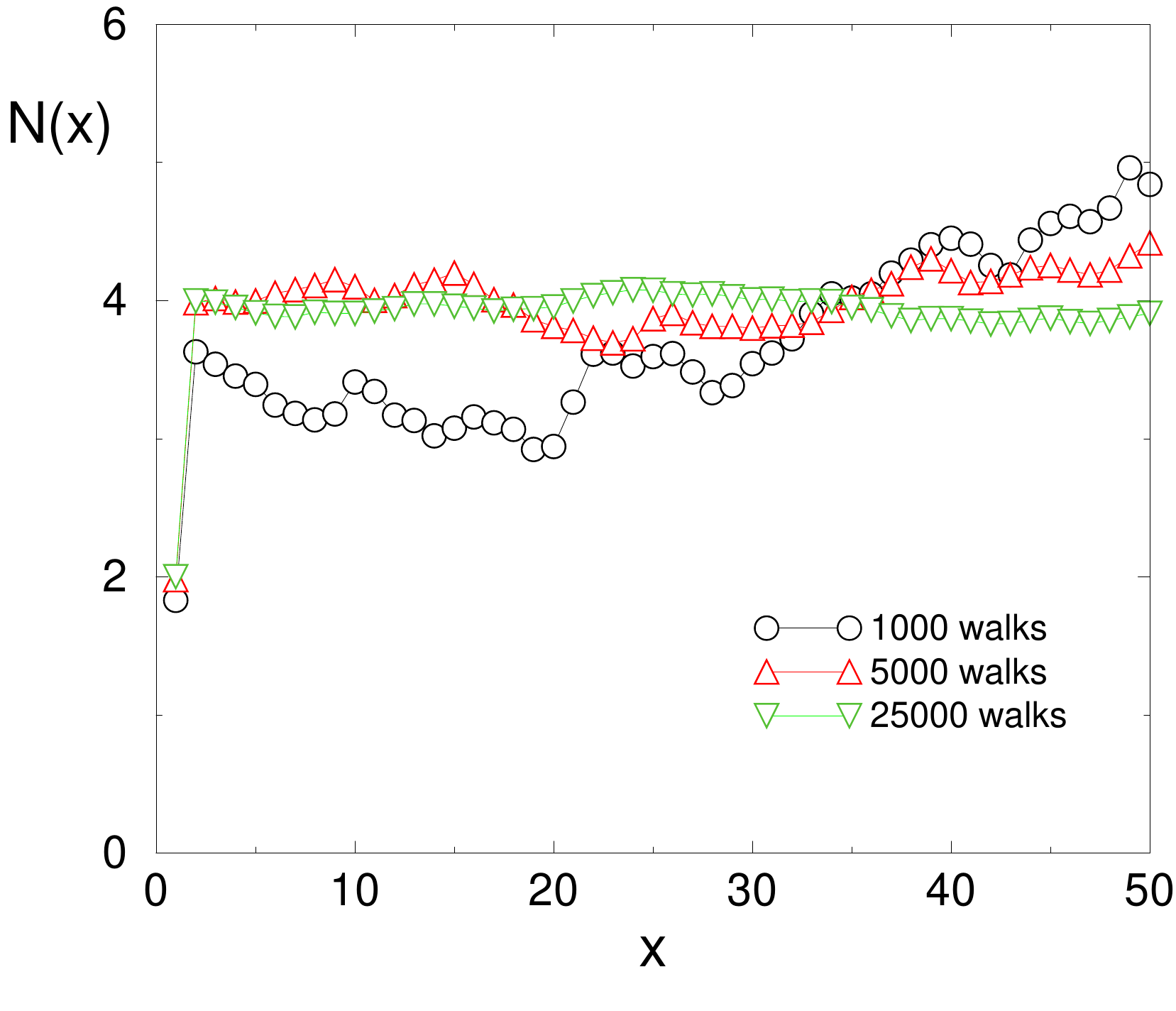}}\qquad
    \subfigure[]{\includegraphics[width=0.425\textwidth]{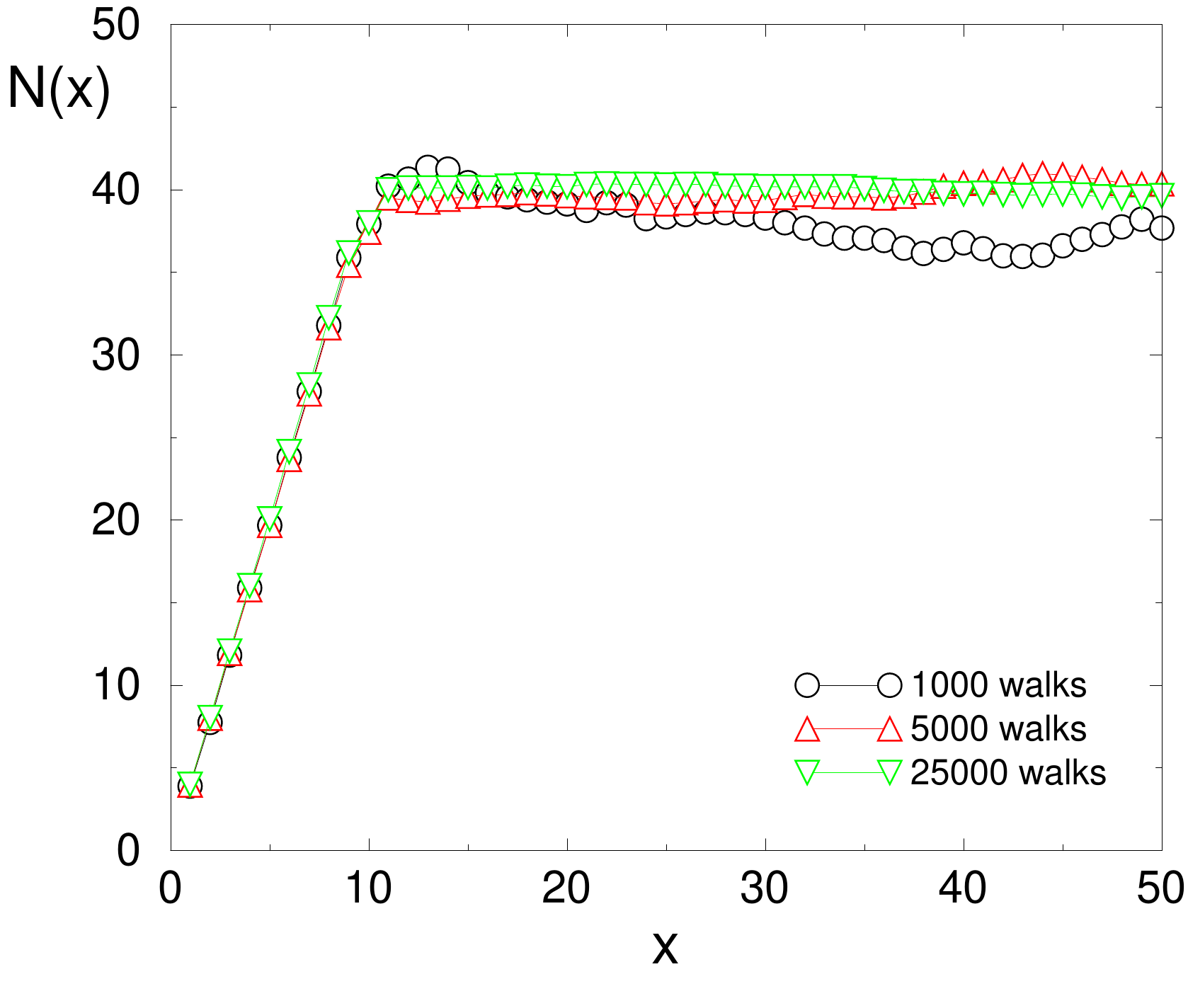}}}
  \centerline{\subfigure[]{\includegraphics[width=0.425\textwidth]{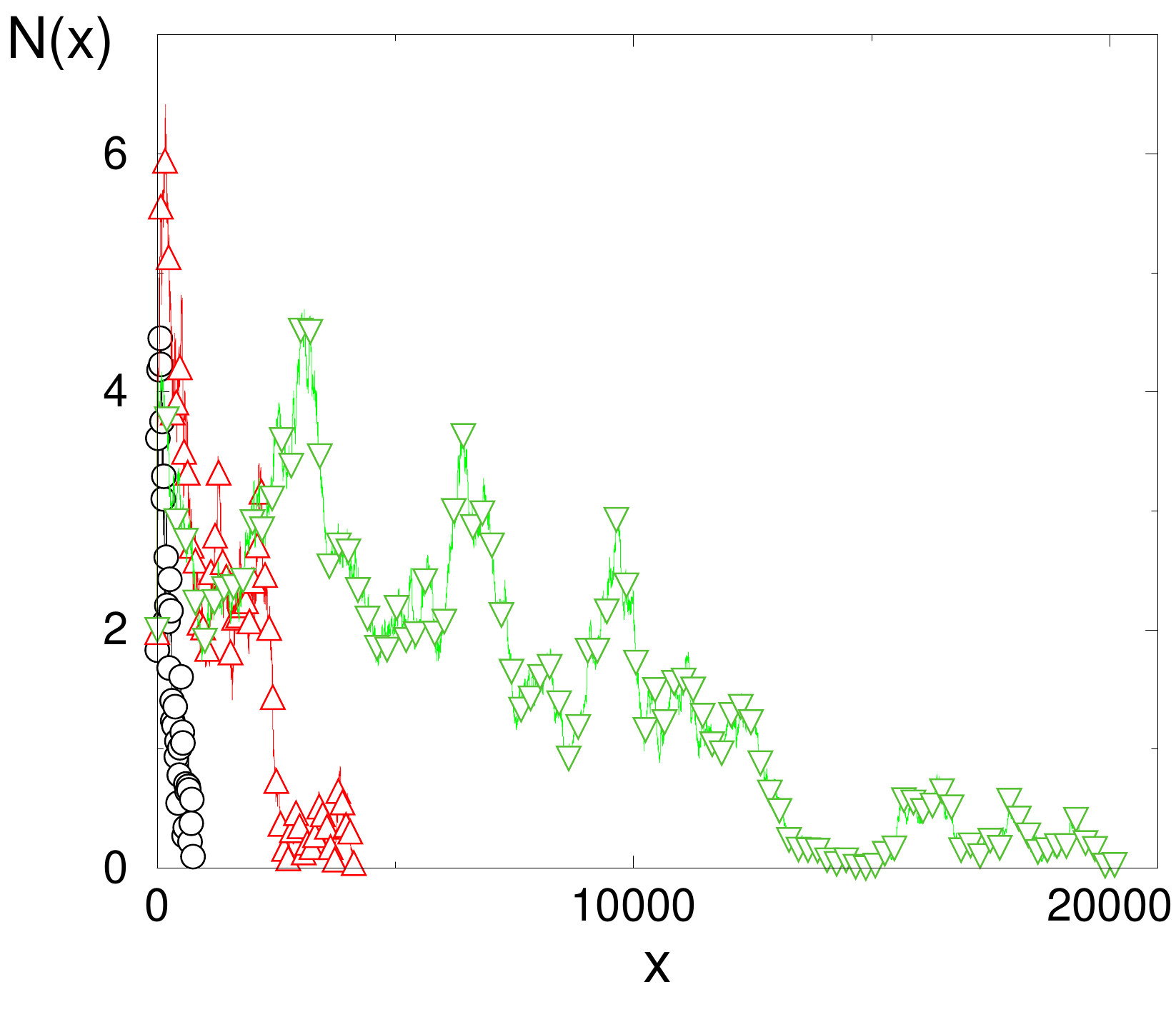}}\qquad
    \subfigure[]{\includegraphics[width=0.425\textwidth]{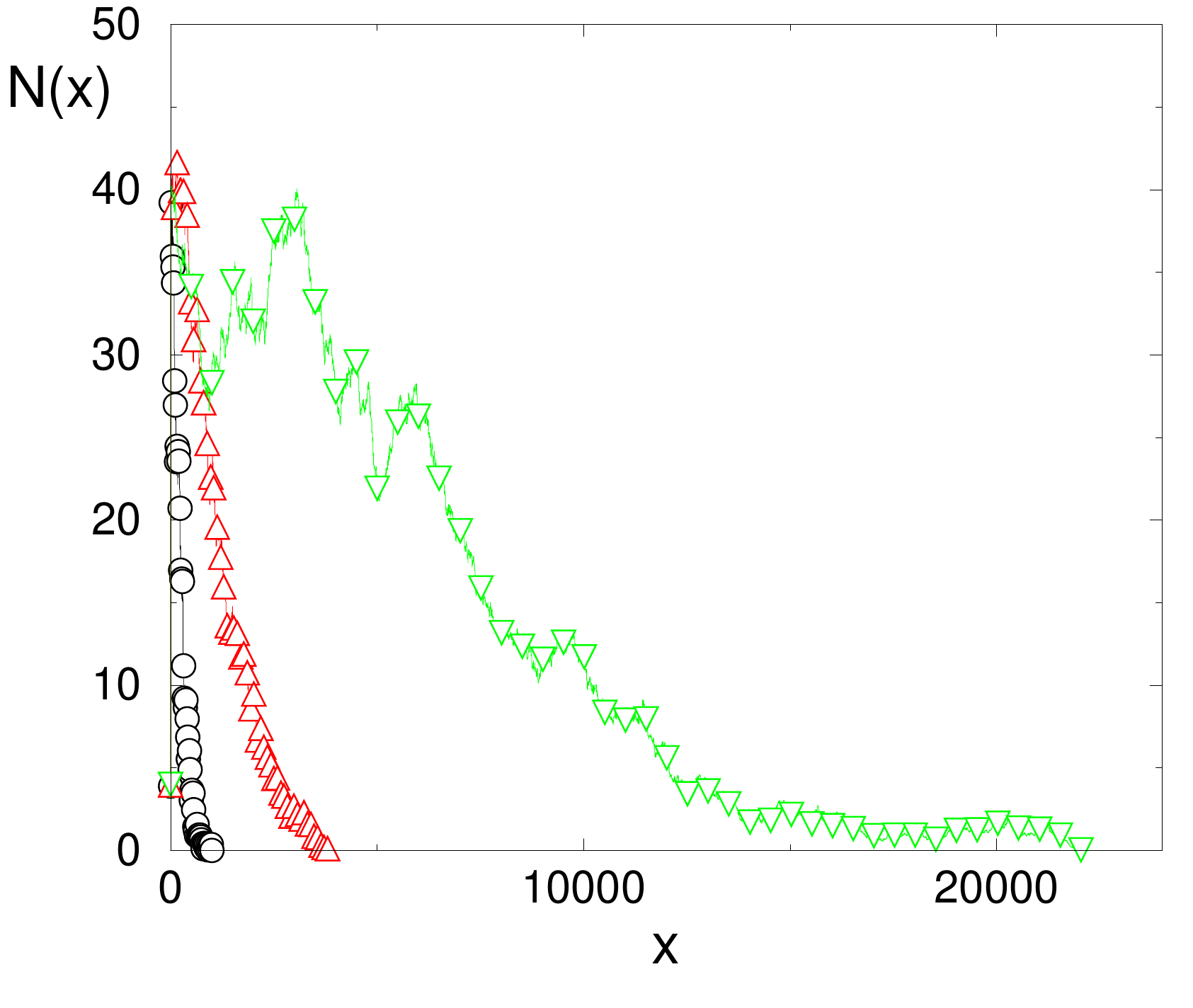}}}
  \caption{Simulation results for $N(x)$, the average number of times that a
    random walk visits $x$ when it starts at: (a) $x_0=1$, and (b) $x_0=10$.
    Here $N(x)$ is the discrete analog of the residence time $T(x)$. (c) \&
    (d): The same data as in (a) and (b) over the full range of $x$. }
\label{Nx}  
\end{figure}

To illustrate this result, we present simulations for a nearest-neighbor
random walk that starts at: (a) $x_0=1$ and (b) at $x_0=10$ in Fig.~\ref{Nx}.
As a function of the number of walks in the ensemble, $T(x)$ slowly converges
to the asymptotic time-independent value in Eq.~\eqref{Nfree}.  A curious
feature of this residence-time data is that it becomes erratic for large $x$,
as shown in Figs.~\ref{Nx}(c) and (d).  We can understand the origin of these
large fluctuations by the following rough argument: for a diffusing particle
that starts at $x_0$, the probability $S(t)$ that it survives until time $t$
is $S(t)=\text{erf}\big(x_0/\sqrt{4Dt}\big)\simeq x_0/\sqrt{4Dt}$ for
$t\to\infty$~\cite{R01}.  For $\mathcal{M}$ random walks, we estimate the
longest lived of them by the extreme-statistics criterion
$S(t_{\rm max})\simeq 1/{\mathcal{M}}$~\cite{G87,KRB10}, which states that
one out of $\mathcal{M}$ walks survives until at least time $t_{\rm max}$.
This criterion gives $t_{\rm max}\simeq (\mathcal{M}x_0)^2/4D$.
Correspondingly, the maximal range reached by an ensemble of $\mathcal{M}$
random walks is, roughly,
$x_{\rm max} \sim \sqrt{Dt_{\rm max}}\sim \mathcal{M}x_0$.

We now use this estimate to determine the large-$x$ fluctuations in
Figs.~\ref{Nx}(c) and (d).  To obtain an accuracy of, say, 10\%, in $N(x)$,
the number of times that the lattice site at $x$ is visited by a random walk,
we need roughly 100 walks to reach this value of $x$.  Since $x_{\rm max}$
scales linearly in the number of realizations, roughly $m=100$ walks will
reach a distance $x=\mathcal{M} x_0/m$.  For example, for 25000 walks
starting at $x_0=1$, roughly 100 of them will reach $x=250$.  Thus up to
$x\approx 250$, the variation in $N(x)$ should be smaller than 10\%, and
beyond this point fluctuations should become progressively more pronounced.
This estimate is consistent with the data of Figs.~\ref{Nx}(c) and (d).

The approach given here can be readily extended to any situation where the
spatial probability distribution can be computed explicitly.  We now
investigate three such cases: (a) biased diffusion, (b) diffusion constrained
to remain in the interval $[0,L]$, and (c) diffusion exterior to an absorbing
sphere in general spatial dimension $d$.

\subsection{Biased diffusion on the semi-infinite line}
\label{subsec:bias}

Suppose that a diffusing particle also experiences a constant bias velocity
$-v$ that systematically pushes the particle towards the origin, so that the
average time for the particle to reach the origin is finite.  For a diffusing
particle that starts at $x_0$, its probability density can be obtained by the
image method~\cite{D82,R01}, and is given by
\begin{align}
\label{Pv}
P(x,t)=\frac{1}{\sqrt{4\pi Dt}}\,\,
  \left[e^{-(x-x_0+vt)^2/4Dt}- e^{-vx_0/D}\,\,e^{-(x+x_0+vt)^2/4Dt}\right]\,.
\end{align}
Notice that the magnitude of the image particle is different from that of the
initial particle, while the velocities of the initial and image particles are
the same.

We again integrate this expression over all time and obtain, for the time
that the particle spends in $[x,x+dx]$ before it dies:
\begin{subequations}
\begin{align}
  \label{Nv}
T(x) &
         =\begin{cases}
           {\displaystyle \frac{dx}{v}\left[1-e^{-vx/D}\right]} & \qquad x<x_0\,,\\[4mm]
          {\displaystyle \frac{dx}{v}\,\,e^{-vx/D}\left[e^{vx_0/D}-1\right]}
          & \qquad x>x_0\,.           \end{cases}
\end{align}
For $v\to 0$, Eq.~\eqref{Nfree} is recovered, while in the opposite
limit of $v\to\infty$, \eqref{Nv} reduces to
\begin{align}
  \label{Nv1}
T(x) \to
         \begin{cases}
           {\displaystyle \frac{dx}{v}} & \qquad x<x_0\,,\\[5mm]
           {\displaystyle  \frac{dx}{v}\,\,e^{-v(x-x_0)/D}} & \qquad x>x_0\,.
           \end{cases}
\end{align}
\end{subequations}
As one might expect, the time spent in $[x,x+dx]$ with $x<x_0$ is just that
of a ballistic particle, while it is exponentially unlikely for the particle
to reach the classically forbidden region $x>x_0$ for large P\'eclet number,
$vx/D$.

\subsection{Diffusion in a finite interval}
\label{subsec:finite}

Suppose that an isotropically diffusing particle is constrained to remain
within the interval $[0,L]$ and is eventually absorbed at $x=0$.  We again
want the time $T(x)$ that the particle spends in $[x,x+dx]$ before it dies.
As in the previous two subsections, we need the spatial probability
distribution for a diffusing particle with absorbing boundary conditions at
$0$ and at $L$.  A straightforward computation of this distribution is
unwieldy, as it involves either an infinite Fourier series or an infinite
number of images.

However, we can avoid this complication by noticing that we only want the
integral of the probability distribution over all time, which corresponds to
its Laplace transform at Laplace variable $s=0$.  The Laplace transform
satisfies $s\widetilde P-\delta(x-x_0)=D\widetilde P_{xx}$, where
$\widetilde P$ denotes the Laplace transform and the subscript denotes
partial differentiation.  For $s=0$, this reduces to the Laplace equation
\begin{align}
  D\widetilde P_{xx}=-\delta(x-x_0)\,.\nonumber
\end{align}
We solve this equation separately in the subdomains $x<x_0$ and $x>x_0$,
impose the boundary conditions, continuity of the solution at $x=x_0$, and
the joining condition $D\big(\widetilde P_x|_{>}-\widetilde P_x|_{<}\big)=-1$
to give, after standard steps,
\begin{equation}
  \widetilde P = \frac{x_<}{D}\Big(1-\frac{x_>}{L}\Big)\,,
\end{equation}
where $\widetilde P_x|_{>}$ is the derivative just to the right of $x_0$ (and
similarly for $\widetilde P_x|_{<}$), and $x_<=\text{min}(x,x_0)$,
$x_>=\text{max}(x,x_0)$.

\begin{figure}[ht]
  \centerline{\subfigure[]{\includegraphics[width=0.33\textwidth]{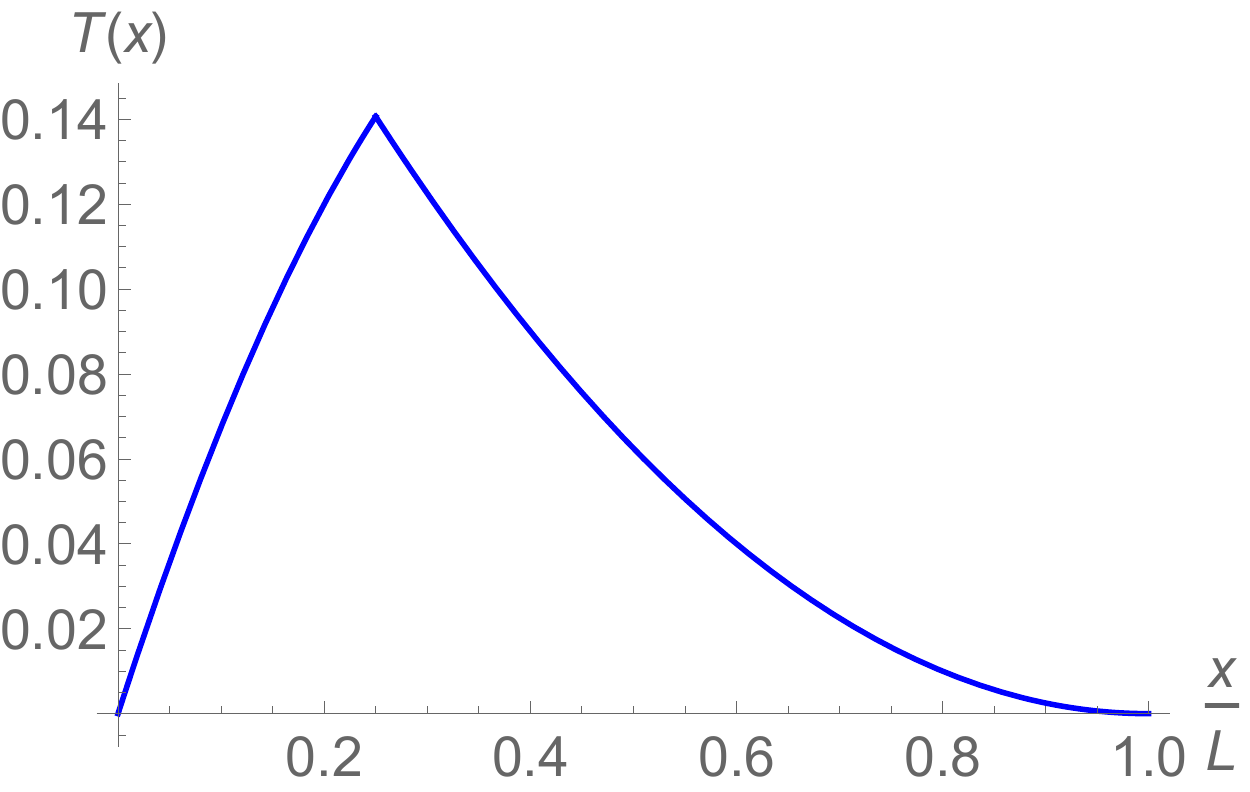}}\quad
    \subfigure[]{\includegraphics[width=0.33\textwidth]{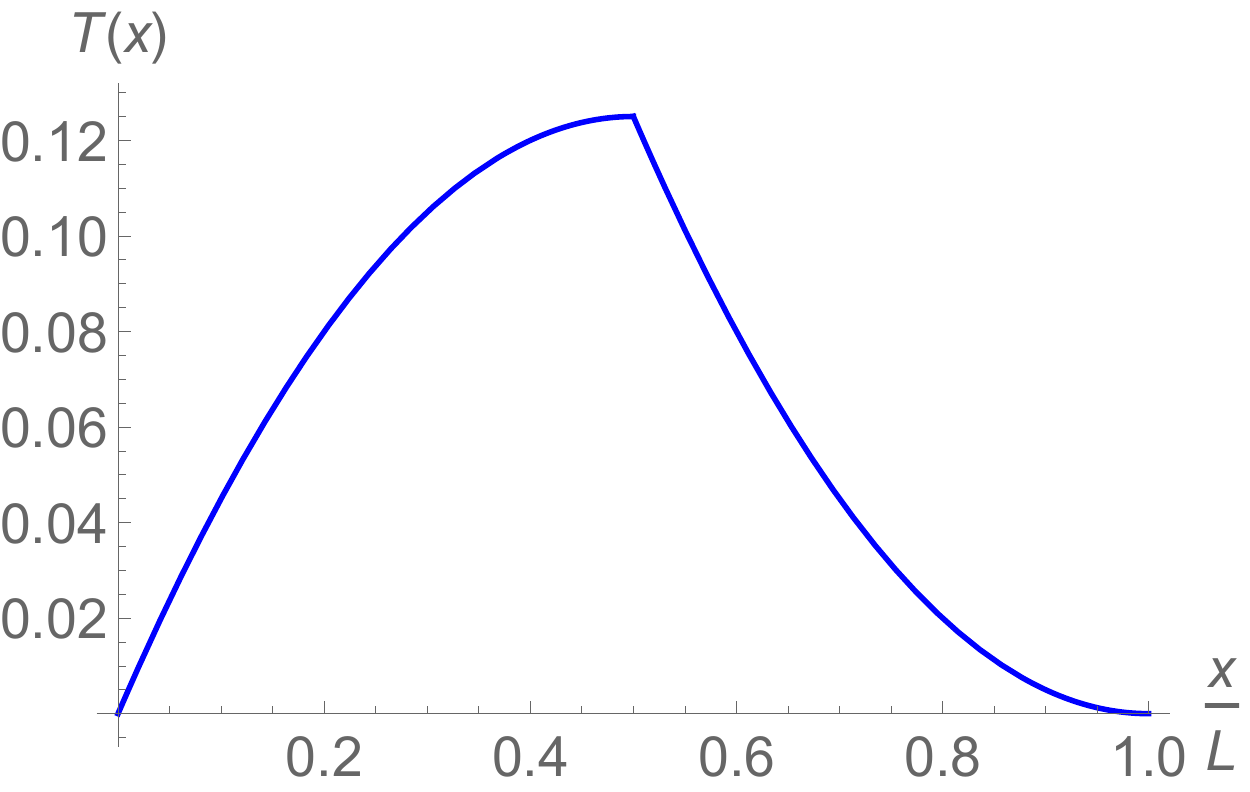}}\quad
\subfigure[]{\includegraphics[width=0.33\textwidth]{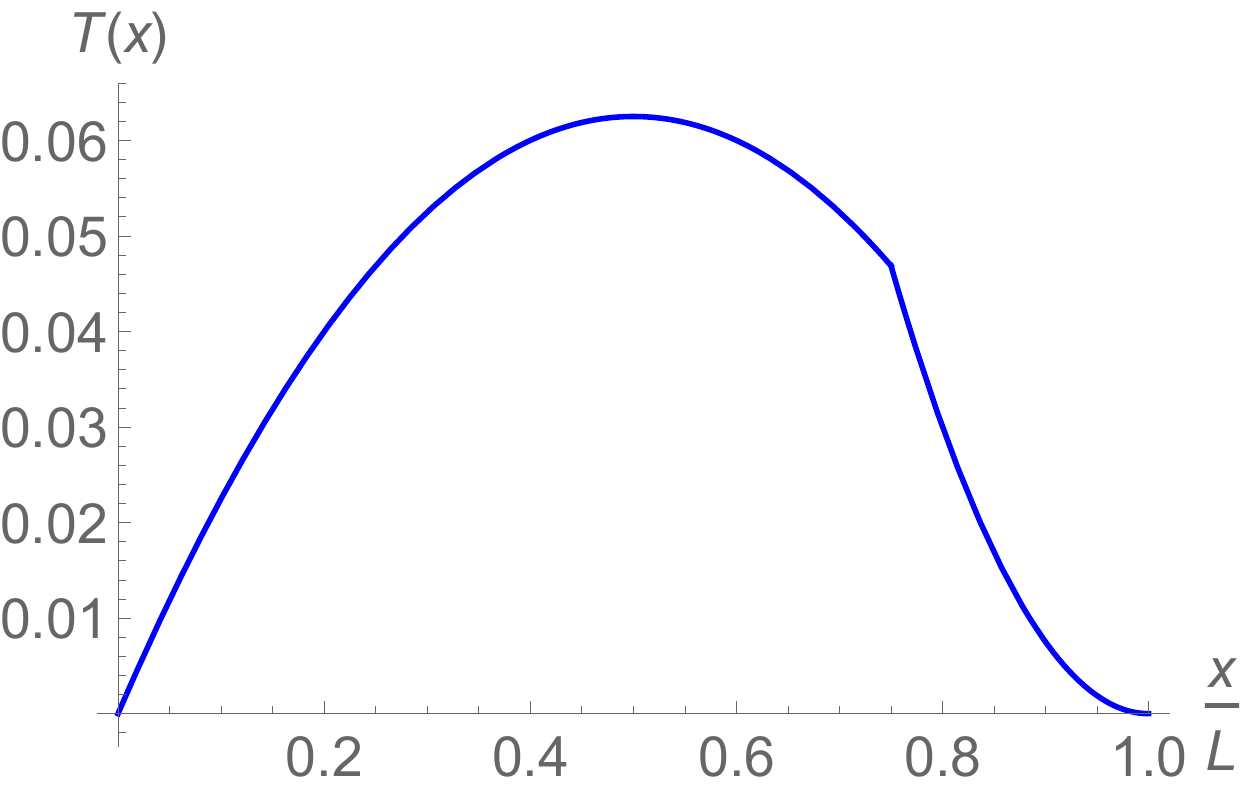}}}
\caption{The residence time density $T(x)/dx$ for a diffusing particle that
  is constrained to remain within the interval $[0,L]$ until it exits at
  $x=0$.  Shown are the cases: (a) $x_0=0.25L$, (b) $x_0=0.5L$ and (c)
  $x_0=0.75L$.}
\label{strip}  
\end{figure}

Finally, to obtain $T(x)$, we need to multiply the above distribution by the
probability that the particle eventually exits the strip at $x=0$, which is
simply $1-\frac{x}{L}$.  Thus we have
\begin{align}
  \label{Nstrip}
 T(x) &=dx\,\int_0^\infty dt\, P(x,t)\left(1-\frac{x}{L}\right)
            =\widetilde P(x,s=0) \left(1-\frac{x}{L}\right)\,dx\nonumber\,,\\[3mm]
          &=  \begin{cases}
    {\displaystyle \frac{x}{D} \Big(1-\frac{x_0}{L}\Big)\Big(1-\frac{x}{L}\Big)\,\, dx} & \qquad x<x_0\,,\\[4mm]
    {\displaystyle \frac{x_0}{D}\Big(1-\frac{x}{L}\Big)^2\,\, dx} & \qquad x>x_0\,.
  \end{cases}
\end{align}
The maximum residence time occurs at $x=x_0$ for $x_0<{L}/{2}$ and then
``sticks'' at $x={L}/{2}$ for $x_0\geq {L}/{2}$, with a cusp always occurring
at $x=x_0$ (Fig.~\ref{strip}).  In the limit $L\rightarrow \infty$, we
recover the result \eqref{Nfree} for diffusion on the semi-infinite line.

\subsection{Diffusion exterior to a sphere in dimension $d>2$}
\label{subsec:gend}

We now determine the residence time for a diffusing particle that wanders in
the region exterior to an absorbing sphere of radius $a$, a geometry that is
the analog of the semi-infinite system in one dimension.  Without loss of
generality, we take the initial condition to be a spherical shell of unit
total probability at radius $r_0$.  We first treat the case of spatial
dimensions $d>2$ and then the special case of $d=2$.

For general $d$, we need to solve the Laplace equation
\begin{align}
  \label{Lap-d}
  D\nabla^2 \widetilde P= -\frac{1}{\Omega_d \,r_0^{d-1}}\,\,\delta(r-r_0)\,,
\end{align}
where $\Omega_d$ is the surface area of a $d$-dimensional unit sphere and
$r_0$ is the radial coordinate of the starting point.  Because of the
spherically symmetric source term, angular coordinates are irrelevant.  We
therefore separately solve $\widetilde P''+\frac{d-1}{r}\widetilde P'=0$ in
the subdomains $a<r<r_0$ and $r_0<r$, and then impose the absorbing boundary
condition at $r=a$ and the joining condition by integrating \eqref{Lap-d}
over an infinitesimal interval that includes $x_0$.  The result of these
standard manipulations is
\begin{align}
  \label{P-gend}
  \widetilde P(r)=
\frac{1}{D(2-d)\Omega_d}\left[\Big(\frac{r_<}{a}\Big)^{2-d}-1\right]r_>^{2-d}\,.
\end{align}

\begin{figure}[ht]
  \centerline{\subfigure[]{\includegraphics[width=0.4\textwidth]{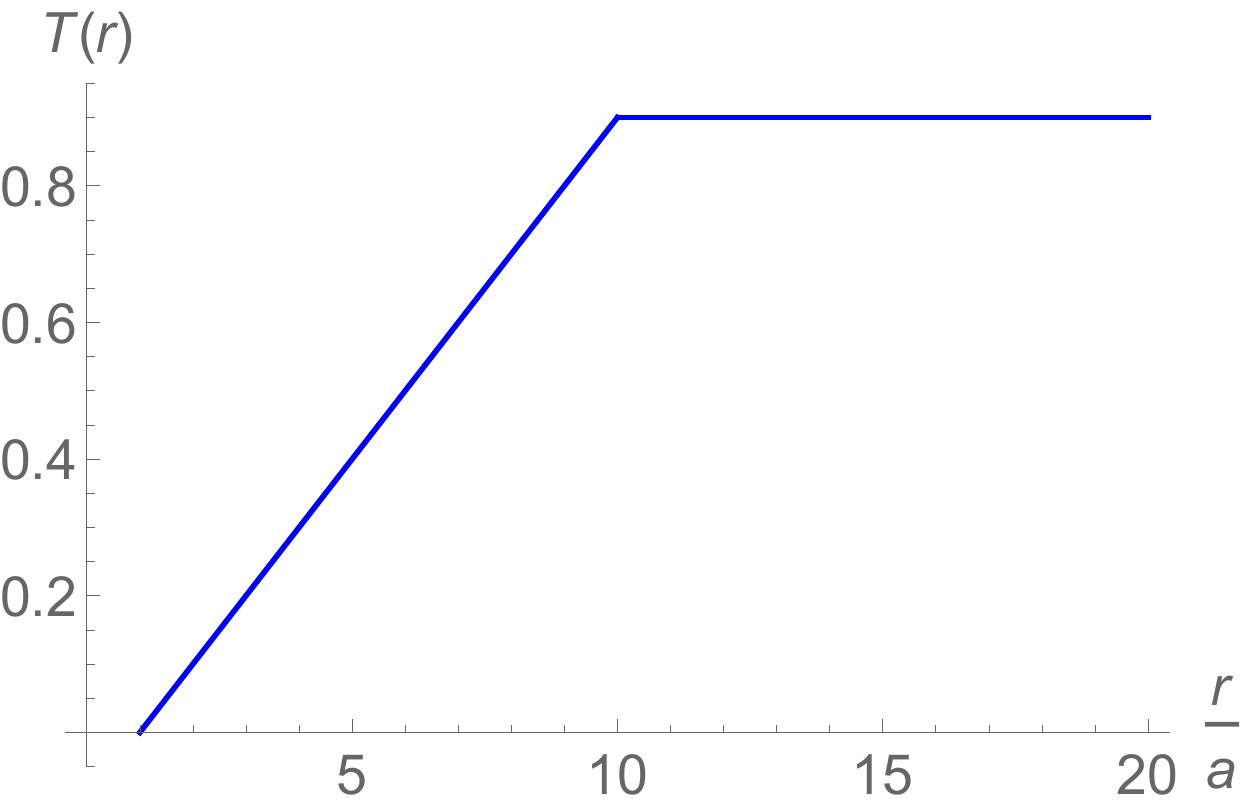}}\qquad\qquad
    \subfigure[]{\includegraphics[width=0.4\textwidth]{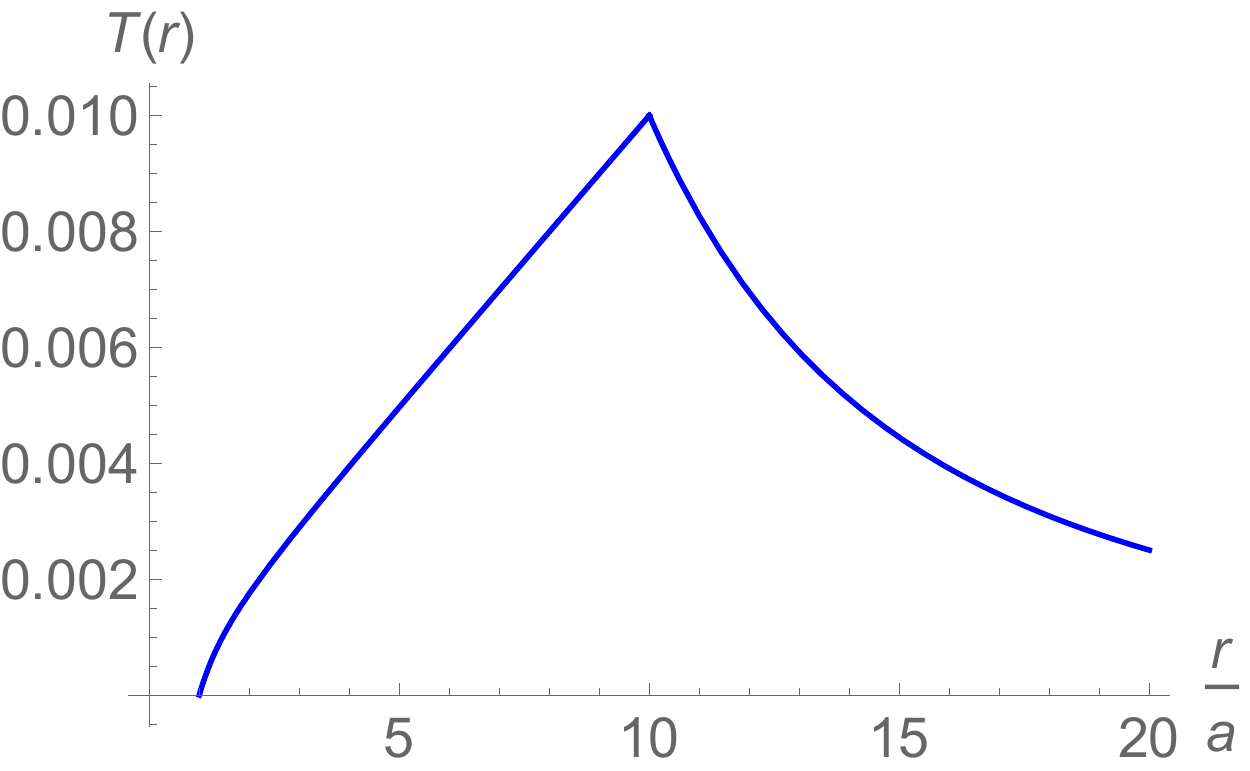}}}
  \caption{The residence time density $T(r)/dr$ for a diffusing particle that
    starts at $r_0=10$ exterior to a sphere of radius 1 in: (a) $d=3$ and (b)
    $d=5$.}
\label{d>2}  
\end{figure}

To obtain $T(r)$, the residence time in a shell of radius $r$ and thickness $dr$, we again need to multiply the above expression by the
probability that the particle eventually hits the sphere, which, for $d>2$,
is simply $(a/r)^{d-2}$~\cite{R01}.  Thus we have
\begin{align}
  \label{Nd}
  T(r) &=  \Omega_ d\,r^{d-1} \widetilde P(r)\, \Big(\frac{a}{r}\Big)^{d-2}dr\,,\nonumber\\
     &=  \begin{cases}
       {\displaystyle \frac{dr}{D(d-2)} \Big(\frac{a}{r}\Big)^{d-2}
         \Big[1-\Big(\frac{a}{r}\Big)^{d-2}\Big]\frac{r^{d-1}}{r_0^{d-2}}}& \qquad r<r_0\,,\\[6mm]
       {\displaystyle \frac{dr}{D(d-2)}\, \Big(\frac{a}{r}\Big)^{d-2}
         \Big[1-\Big(\frac{a}{r_0}\Big)^{d-2}\Big]\,r} & \qquad r>r_0\,.
  \end{cases}
\end{align}
Two representative results are shown in Fig.~\ref{d>2}.  For large spatial
dimension, a particle that eventually hits the sphere of radius $a$ must do
so quickly.  Thus the residence time in the domain $r>r_0$ must necessarily
be small, as shown in Fig.~\ref{d>2}(b) for $d=5$.

In spatial dimension $d=2$, the result analogous to Eq.~\eqref{P-gend} is
\begin{align}
  \widetilde P(r) =  \frac{1}{2\pi D}\,\ln\frac{r_<}{a}\,.
\end{align}
In distinction to the cases of $d=1$ and $d>2$, $\widetilde P(r)$ is constant
for $r>r_0$.  Since a diffusing particle always reaches the absorbing sphere
in $d=2$, we immediately have $T(r) = 2\pi r \widetilde P(r)\, dr$.

\section{Visitation by a Discrete Random Walk}
\label{sec:rw}

We now investigate the corresponding residence time for a symmetric random
walk in the semi-infinite one-dimensional domain $[0,\infty]$.  The walk
starts at lattice site $x_0$ and is absorbed when it first reaches $x=0$.
The analog of the residence time is $N(x)$, the number of times that the
random walk visits site $x$ (excluding the initial visit if $x=x_0$) before
the walk dies.  We use the generating function approach to derive this
quantity for the case of $x_0=1$.

\subsection{Average number of  revisits to $x=1$}
\label{subsec:x=1}

\begin{figure}[ht]
  \centerline{\includegraphics[width=0.6\textwidth]{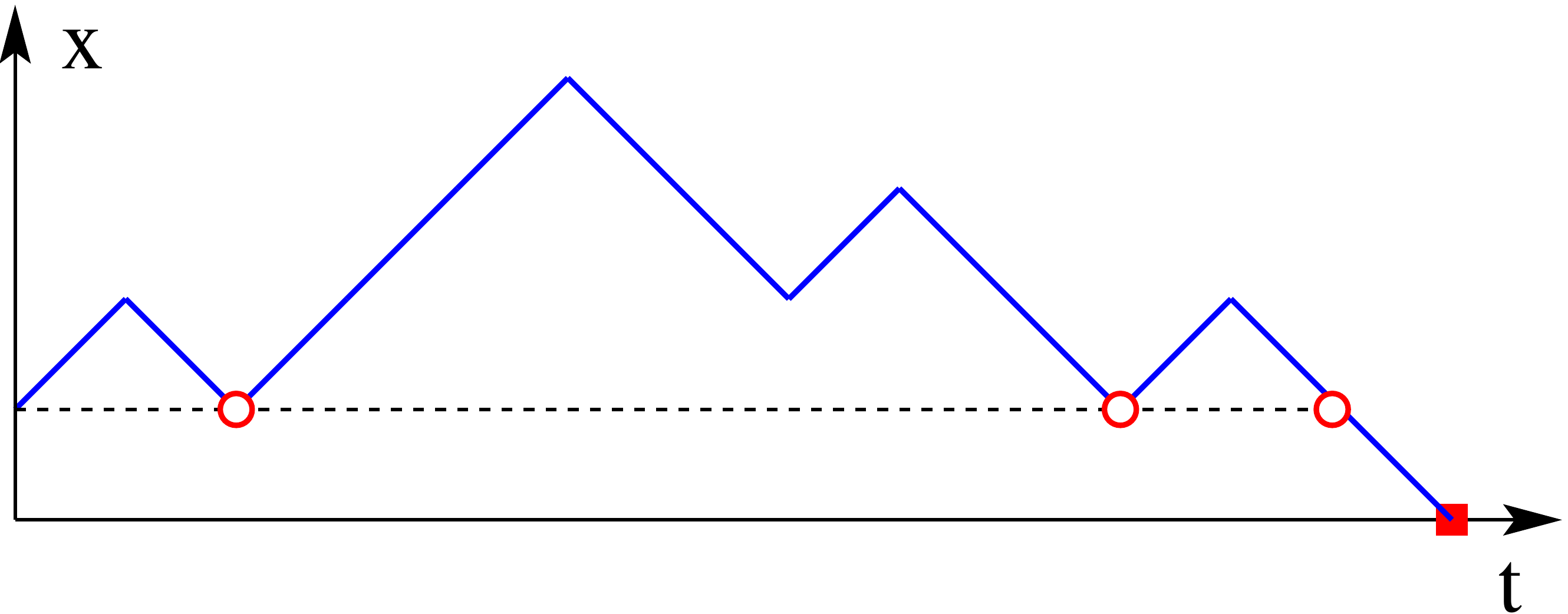}}
  \caption{Space-time trajectory of a one-dimensional random walk of
    $2n+1=13$ steps that starts at $x=1$ and makes 3 revisits to $x=1$ (red
    circles) before being absorbed at $x=0$ (square).  This path contributes
    to $A(6,3)$ (see Eq.~\eqref{Ank}).}
\label{A6-3}  
\end{figure}

For a random walk that starts at $x_0=1$ and is absorbed at $x=0$, the number
of steps in the walk is necessarily odd.  For convenience, we write this
number as $2n+1$, with $n$ an arbitrary non-negative integer.  We define
$A(n,k)$ as the number of random-walk paths that start at $x_0=1$, take the
first step to the right (thus upward in the space-time representation of
Fig.~\ref{A6-3}), and make $k$ revisits to $x=1$, before dying at the
$(2n+1)$st step.  The number of such paths was found in~\cite{BRB17} and is
given by
\begin{align}
\label{Ank}
A(n,k)= \frac{k\,(2n-k-1)!}{(n-k)!\,n!}\,,
\end{align}
which happens to be directly related to the triangular Catalan
numbers~\cite{B96,LO16}.  To compute the average number of revisits to $x=1$,
we will need $\mathbf{P}(k\,|\,n)$, the conditional probability for a path to
make exactly $k$ revisits to $x=1$ before dying at step $2n+1$.  This
probability is
\begin{align}
  \label{Pkn}
  \mathbf{P}(k\,|\,n)={A(n,k)}/{C_n}\,,
\end{align}
where $C_n= \frac{1}{n+1}\binom{2n}{n}$ is the $n$th Catalan
number~\cite{S15}, which counts the total number of random walks of $2n$
steps that start at $x=0$, remain in the region $x\geq 0$, and return to $x=0$ at step $2n$.  For what
follows, we will also need
\begin{align}
  P(n) = C_n/2^{2n}\,,
\end{align}
the probability that a random walk first returns to its starting point at
step $2(n+1)$.  Note the shift $n\to n+1$ to ensure that the walk always
remains above $x=0$.

From Eqs.~\eqref{Ank} and \eqref{Pkn}, we have
\begin{align}
  \label{Pnk}
  \mathbf{P}(k\,|\,n)= \frac{k\,(2n-k-1)!\,(n+1)!}{(n-k)!\, (2n)!}=k\,{\binom{n+1}{k+1}}\bigg/{\binom{2n}{k+1}}\,.
\end{align}
Thus the number of revisits to $x=1$, averaged over all walks of $2n+1$ steps
is given by
\begin{align}
\langle k\rangle_n = \sum_{k=1}^{\infty}\,k\,\mathbf{P}(k\,|\,n)\,.
\end{align}
Using expression \eqref{Pnk} for $\mathbf{P}(k\,|\,n)$ in the above average, we obtain the
remarkably simple result
\begin{align}
  \label{3n}
  \langle k\rangle_n =
  \sum_{k=1}^{\infty}\,k^2\,{\binom{n+1}{k+1}}\bigg/{\binom{2n}{k+1}}
  =  \frac{3n}{n+2}\,.
\end{align}
For long paths of $2n+1$ steps, there are, on average, $3$ revisits to $x=1$,
after which the walk immediately dies.

We now determine the number of revisits to $x=1$ upon also averaging over all
$n$.  This double average is
\begin{align}
  \label{kavav}
 \langle\!\langle k\rangle\!\rangle= \sum_{n,k\geq 1}k\, \mathcal{P}(n,k)\,.
\end{align}
Here $\mathcal{P}(n,k)$ is the joint probability that the walk first reaches
$x=0$ at step $2n+1$, \textit{and} the walk makes $k$ revisits to~$x=1$
within $2n+1$ steps.  This joint probability is
\begin{align}
  \mathcal{P}(n,k) = \mathbf{P}(k\,|\,n)\,P(n)
                   = \frac{A(n,k)}{C_n} \,\frac{C_n}{2^{2n}}
                   = \frac{A(n,k)}{2^{2n}}\,.
\end{align}                     
The average in \eqref{kavav} may now be expressed in terms of the generating
function for $\mathcal{P}(n,k)$:
\begin{align}
\label{g}
  g(x,y)=  \sum_{n,k\geq 1}\, \mathcal{P}(n,k)\,x^n\,y^k
  &=\sum_{n,k\geq 1}\, \frac{1}{2^{2n}}\, A(n,k) \,x^n\,y^k\,,\nonumber\\
   &=\sum_{n,k\geq 1}\, \frac{1}{2^{2n}}\, \frac{(2n-k-1)! \,k}{(n-k)!\,n!} \,x^n\,y^k\,,\nonumber\\
&= \frac{xy}{2-xy+2\sqrt{1-x}}\,,
\end{align}
Which was derived in~\cite{BRB17} (see also~\cite{S99}).  In terms of the
generating function, we immediately obtain the remarkably simple result
\begin{align}
\label{2}
  \langle\!\langle k\rangle\!\rangle=
   \sum_{n,k\geq 1}\, k\, \mathcal{P}(n,k)\,x^n\,y^k\,\Big|_{(1,1)}
  = y\,\,\frac{\partial g}{\partial y}\,\bigg|_{(1,1)}\!\! =2\,.
\end{align}
There are, on average, 2 revisits to $x=1$ in the ensemble of all random
walks that start at $x=1$, take their first step to the right, and are
eventually absorbed at $x=0$.

We can extend Eq.~\eqref{3n} to higher integer moments of the average number
of revisits to $x=1$ for walks of $2n+1$ steps.  The first few of these
fixed-$n$ moments are:
\begin{subequations}
\begin{align}
  \langle k^2\rangle_n &= \frac{n(13n-1)}{(2+n)(3+n)}\,,\nonumber\\
  \langle k^3\rangle_n &= \frac{15n^2(5n-1)}{(2+n)(3+n)(4+n)}\,,\nonumber\\[3mm]
  \langle k^4\rangle_n &= \frac{541n^2-196n^2+11n^2+4n}{(2+n)(3+n)(4+n)(5+n)}\,,
\end{align}
etc.  We can similarly compute the higher integer moments of the number of
revisits to $x=1$, averaged over all walk lengths, and the first few are:
\begin{align}
  \langle\!\langle k^2\rangle\!\rangle = 6\qquad
  \langle\!\langle k^3\rangle\!\rangle = 26\qquad
  \langle\!\langle k^4\rangle\!\rangle = 150\qquad
  \langle\!\langle k^5\rangle\!\rangle = 1082\,,
\end{align}
\end{subequations}
etc.  Parenthetically, these numbers are also sequence A000629 in the On-Line
Encyclopedia of Integer Sequences~\cite{IS}

In the next section, we will also need the generating function when the first
step of the walk can equiprobably be to the right or to the left.  This leads
to the possibility that the total number of steps $2n+1=1$, i.e., $n=0$, for
which the number of revisits to $1$ equals zero.  The generating function for
the joint probability $\mathcal{P}(n,k)$ for this ensemble of random walks therefore is
\begin{align}
  \label{G}
  G(x,y)=\sum_{n,k\geq 0}\,\mathcal{P}(n,k)\,x^n\,y^k
  &= \tfrac{1}{2}\big[1+g(x,y)\big]\,,\nonumber\\
&= \frac{1}{2}\left(1+\frac{xy}{2-xy+2\sqrt{1-x}}\right)\,,
\end{align}
where the term 1 in the parenthesis comes from the walk that initially steps
to the left and is immediately absorbed.  Notice that
$y\frac{\partial G}{\partial y}\big|_{(1,1)}=1$, which is consistent with
\eqref{g}: half of all paths die immediately upon the first step, and thus
never return to $1$, while the other half return twice, on average, as
derived in \eqref{2}.

\subsection{Average number of visits to $x=2$}

We now extend the above approach to a walk that starts at $x=1$ and is
constrained to take its first step to the right, to determine the number of
visits to $x=2$.  For this purpose, we define three random variables that
characterize this set of walks:
\begin{itemize}
\item $2n+1$, the total number of steps in the walk when it dies;
\item $k$, the number of visits to $x=2$ (\emph{including} the first visit);
\item $\ell$, the number of excursions that lie above the level $x=1$.
\end{itemize}
Since an excursion is a path that lies between two successive returns to
$x=1$ (and thus always remains above $x=1$), the minimal length excursion is
the path $1\to 2\to 1$.

\begin{figure}[ht]
  \centerline{\includegraphics[width=0.6\textwidth]{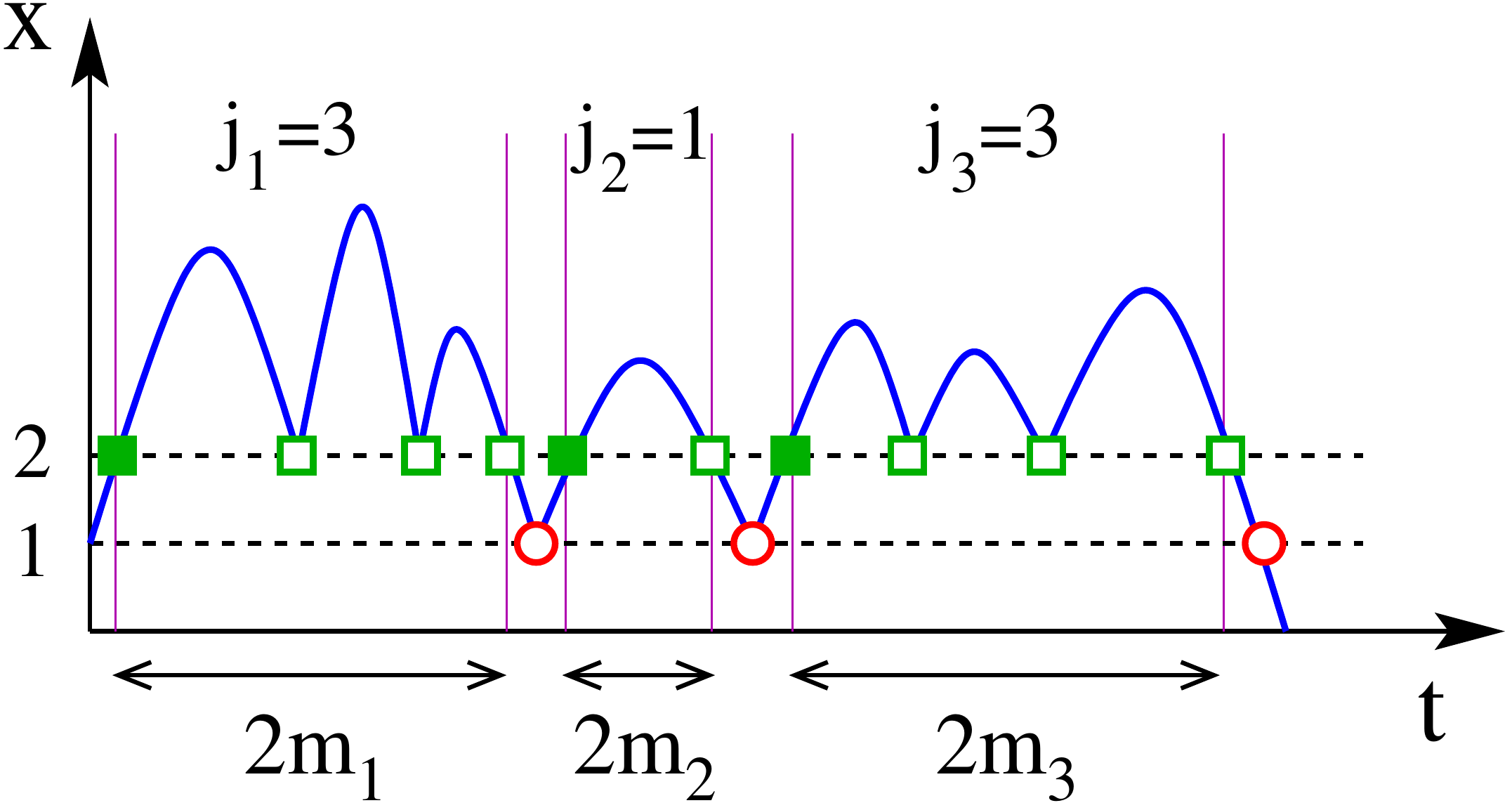}}
  \caption{Schematic space-time trajectory of a random walk that starts at
    $x=1$ and has 3 revisits to $x=1$ (red circles) and $k=10$ visits to
    $x=2$ (green squares). There are $\ell=3$ excursions above $x=1$.
    Immediately after a revisit to $x=1$, the next revisit to $x=2$ is shown
    as a solid green square. }
\label{x2}  
\end{figure}

We want the ensemble average of the number of visits to $x=2$.  To
facilitate this calculation, it is useful to define the three-variable
generating function
\begin{align}
\label{genG}
  \mathcal{G}(x,y,z)=\sum_{n\geq 1}\,\sum_{1\leq \ell\leq n}\, \sum_{\ell\leq  k\leq n}\, \mathcal{P}\left(n,k,\ell\right)\,x^n\,y^k\,z^\ell\,,
\end{align}
which encodes all paths according to $(n,k,\ell)$.  We also label each
successive excursion of the path above $x=1$ by the index
$1\leq i \leq \ell$, and we introduce the variables $2m_i$ and $j_i$,
respectively, for the number of steps in the $i$th such excursion, and the
number of returns to $x=2$ in this excursion (Fig.~\ref{x2}).  As shown in
this figure, $2m_i$ counts the number of steps that lie above $x=2$.  Thus
for an excursion that goes from $x=1$ to $x=2$ and immediately returns to
$x=1$, $m_i=0$.  In addition, $j_i$ counts the number of \textit{revisits} to
$2$, so that the total number of visits to $x=2$ in the $i$th excursion above
$x=1$ is $j_i+1$.  The variables $j_i$, $m_i$, and $\ell$ must satisfy the
geometric constraints (see Fig.~\ref{x2}):
\begin{align}
  \label{geom}
\begin{split}
&j_1+j_2+\dots+j_\ell+\ell = k\,,\\
&m_1+m_2+\dots+m_l +\ell= n\,,\\
&j_i \leq m_i\,.
\end{split}
\end{align}

Using these definitions, the three-variable generating function
$\mathcal{G}(x,y,z)$ can be functionally expressed in terms of $G(x,y)$
defined in Eq.~\eqref{G} as (see the Appendix for details of this derivation)
\begin{align}
  \label{calG}
  \mathcal{G}(x,y,z)
=\sum_{\ell\geq 1}\,P(\ell)\,\left[x\,y\,z\,G(x,y)\right]^\ell\,,
\end{align}
where $P(\ell)$ is the probability that there are $\ell$ excursions above
$x=1$ averaged over walks of any length, which is also the distribution of
the number of returns to~$x=1$.

One may compute $P(\ell)$ as the marginal of the joint distribution of the
number of steps and the number of excursions:
\begin{align}
  \label{P-ell}
P(\ell)=\sum_{n\geq 1} \mathcal{P}(n,\ell) = \sum_{n\geq 1}\frac{1}{2^{2n}}\, A(n,\ell)\,.
\end{align}
The above sum starts at $n=1$ because we are imposing the condition that the
first step of the walk is to the right.  Consequently the three-variable
generating function in \eqref{calG} will ultimately be expressed in terms of
the restricted generating function $g$.  Substituting Eq.~\eqref{P-ell} in
\eqref{calG} and comparing the resulting formula with the first line of
\eqref{g}, we obtain
\begin{align}
\mathcal{G}(x,y,z) &=\sum_{\ell\geq 1}\,P(\ell)\,\big[x\,y\,z\,G(x,y)\big]^\ell\nonumber\\
       &= \sum_{\ell\geq 1}\,
    \sum_{n\geq 1} \frac{1}{2^{2n}}\, A(n,\ell)\, \big[x\,y\,z\,G(x,y)\big]^\ell\nonumber\\[1mm]
&= g\big(1,x\,y\,z\,G(x,y)\big)\,.
\end{align}
It is now straightforward to calculate $\langle k \rangle$.  From the
definition of the generating function \eqref{genG}, we have
\begin{align}
  \langle k\rangle &= y\,\frac{\partial \mathcal{G}}{\partial y}\Big|_{x=y=z=1}\,,\nonumber\\
                   &= y\,\frac{\partial g}{\partial y}\Big|_{x=y=1}\ \left[G(1,1)
                     +y\frac{\partial G}{\partial y}\Big|_{x=y=1}\right]\,,\nonumber\\
                   &= 2\times\left(1+1\right)\,,\nonumber \\
                   &= 4\,.
\end{align}
A random walk thus visits $x=2$ twice as often as $x=1$, as already predicted
by the continuum solution \eqref{Nfree}.

\subsection{Average number of visits to $x> 2$}

The ensemble average of the number of visits to a given level $x> 2$ may be
readily computed by induction.  We start by calculating the average number of
visits to $x=3$, and it will become apparent that this approach applies for
any $x>2$.  Each time a random walk reaches $x=2$, there are two
possibilities at the next step: the walk may step forward to $x=3$ or step
back to $x=1$.  Let us first assume that the walk goes to $x=3$, which occurs
with probability $\frac{1}{2}$.  Each time this event occurs, we now ask:
what is the average number of visits to $x=3$ (including this first visit)
before the walk returns to $x=2$?

With probability $\frac{1}{2}$, the walk may immediately return to $x=2$, in
which case, there is one visit to $x=3$.  On the other hand, if the walk
steps to $x=4$, we have the same situation as that discussed in
Sec.~\ref{subsec:x=1}.  Namely, if we view $x=3$ as the starting point, we
know that there are 2 revisits to $x=3$ and thus 3 visits to $x=3$, on
average, before the walk steps back to $x=2$.  Thus each time $x=3$ is
reached, there are
\begin{align}
\Big(  \frac{1}{2} \times 1 \Big)+ \Big( \frac{1}{2} \times 3\Big) = 2\nonumber
\end{align}
two visits, on average, to $x=3$.

For a walk that reaches $x=2$, the average number of visits to $x=3$ for this
visit to $x=2$ therefore is
\begin{align}
   \Big( \frac{1}{2}\times 0 \Big) + \Big(\frac{1}{2} \times 2 \Big)= 1\,.\nonumber
\end{align}
The first term corresponds to the contribution from a walk that steps from
$x=2$ to $x=1$ without hitting $x=3$, and the second term is the contribution
when the walk steps from $x=2$ to $x=3$.

To summarize, each time the walk visits $x=2$, there is, on average, one
visit to $x=3$, before the walk is at $x=2$ again.  Clearly, this reasoning
that determines the number of visits to $x+1$ for each visit to $x$ applies
inductively for any level $x>2$.  Thus we conclude that the average number of
times that a random walk visits a given level $x>2$, equals 4, in agreement
with the simulation results in Fig.~\ref{Nx}(a).  Clearly, our argument also
applies for any starting point of the walk $x_0$, as long as we restrict to
coordinates with $x>x_0+1$.

\section{Time of the First Revisit}
\label{sec:first}

In addition to the \emph{number} of revisits to $x=1$ by a random walk
excursion that starts at $x=1$ and is eventually absorbed, we are interested
in the \emph{time} at which the first revisit occurs.  This time
characterizes the shape of the space-time trajectory of a random walk.  Since
the walk starts at $x=1$ and ends at $x=0$, its space-time shape is
essentially that of a Brownian excursion ---~a Brownian trajectory that starts
at $x=0$, remains above $x=0$ for all $0<t<T$, and returns to $x=0$ for the
first time at $t=T$.  The average shape of a Brownian excursion has been
shown to be semi-circular~\cite{BCC03,CBC04}.  From this shape, we might
anticipate that the first revisit to $x=1$ is unlikely to occur for $t$ near
$T/2$ because such a revisit involves a large fluctuation from the average
trajectory.  Instead, it seems more likely that the first revisit to $x=1$
will occur either near the beginning or the end of the excursion, a feature
that evokes the famous arcsine laws~\cite{F64,MP10}.  We now show that this
expectation is correct.

\subsection{Time of first return to $1$ for fixed walk length $n$}

\begin{figure}[ht]
  \centerline{\includegraphics[width=0.6\textwidth]{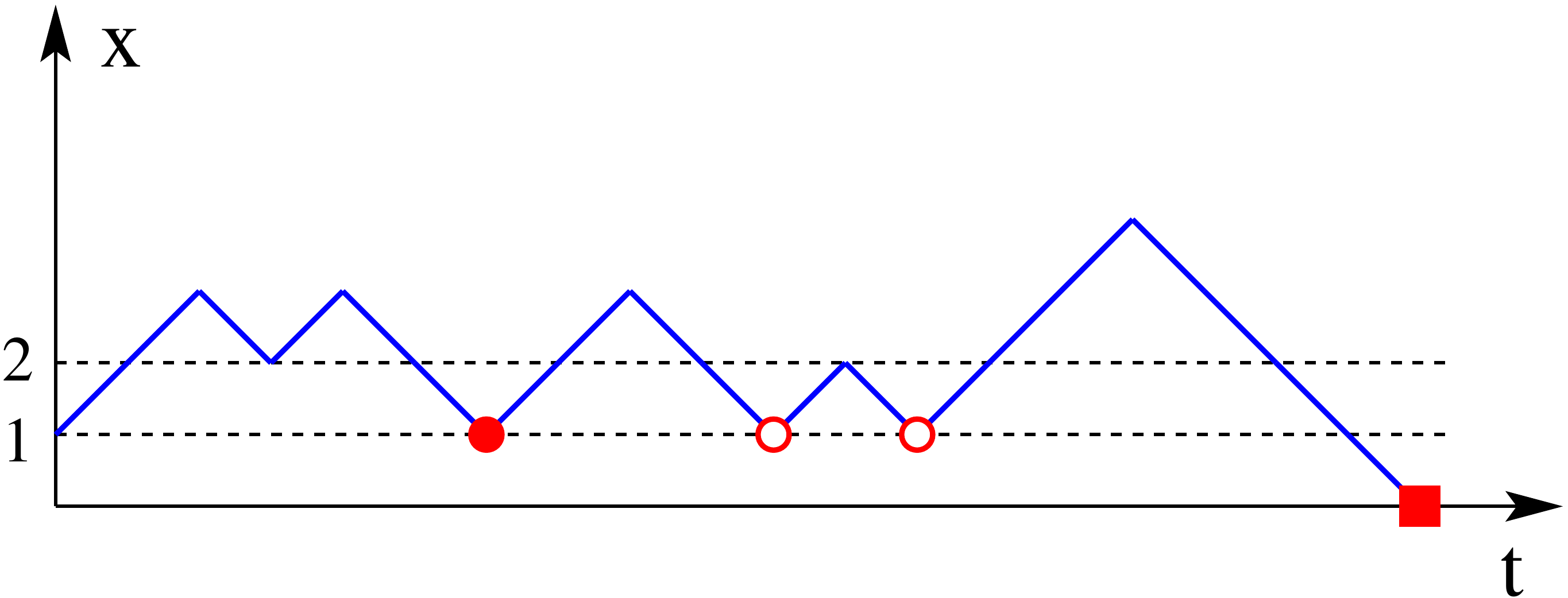}}
  \caption{Space-time trajectory of a one-dimensional random walk of
    $2n+1=19$ steps that starts at $x=1$ and first revisits $x=1$ at step
    $2m=6$ (solid circle).  Subsequent revisits to $x=1$ are indicated by
    open circles and the walk is absorbed when it first reaches $x=0$
    (square).}
\label{1st-revisit}  
\end{figure}

Consider a random walk that starts at $x=1$, takes its first step to the
right, and is absorbed at $x=0$ after $T=2n+1$ steps.  What is the
probability $P(2m\,|\,T)$ that such a walk revisits $x=1$ \emph{for the first
  time} at step $\tau_1=2m$?  Since the walk necessarily revisits $x=1$ at
step $2n$ by definition, and the walk could revisit $x=1$ immediately after
$2$ steps, $m$ satisfies the constraint $1\leq m \leq n$.  The number of
walks that revisit $x=1$ after $2m$ steps may be obtained by decomposing the
full path into two constituents (Fig.~\ref{1st-revisit}):
\begin{itemize}
\item Excursions of $(2m-2)$ steps that wander in the domain $x\geq 2$ ---~the number of
  such paths is $C_{m-1}$;
\item Excursions of $(2n-2m)$ steps that wander in the domain $x\geq 1$ ---~the
  number of such paths is $C_{n-m}$.
\end{itemize}
The first part accounts for the first return to $x=1$ at step $2m$ and the
second part accounts for the remaining path of $2n-2m$ steps.

\begin{figure}[ht]
  \centerline{
    \subfigure[]{\includegraphics[width=0.47\textwidth]{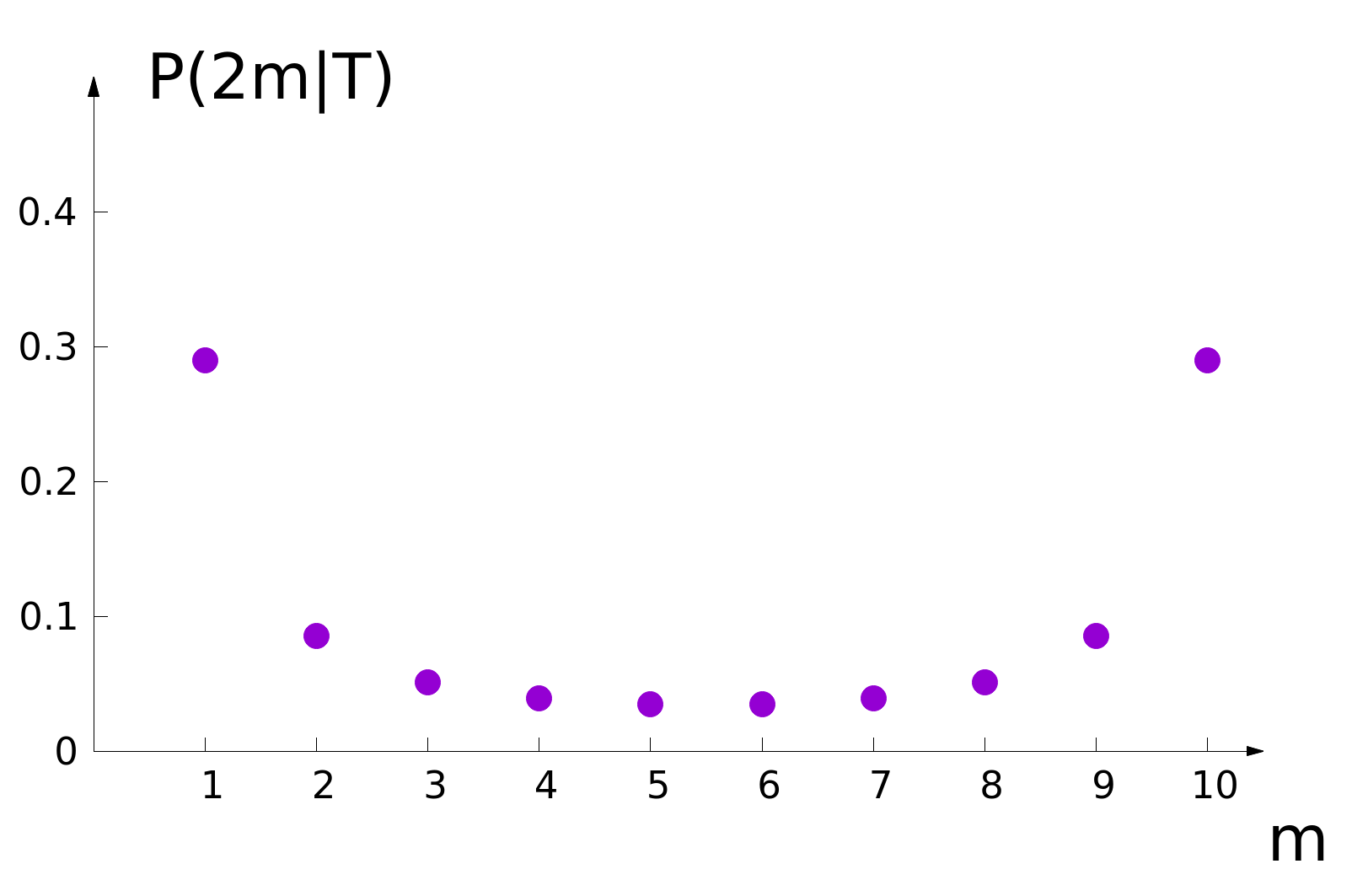}}\quad
    \subfigure[]{\includegraphics[width=0.47\textwidth]{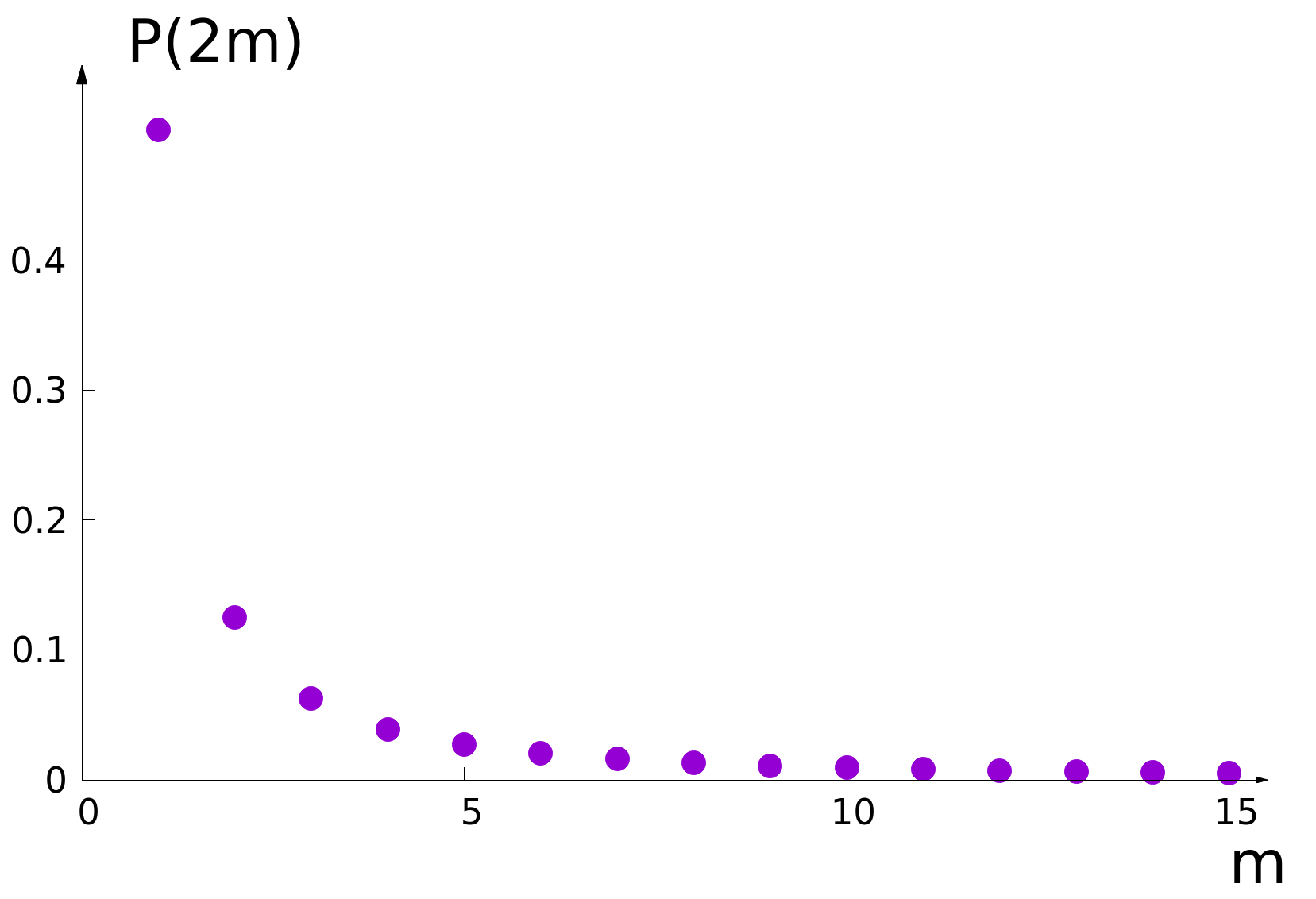}}}
  \caption{(a) Conditional distribution
    $P\left(\tau_1=2m \,| \,T=2n+1\right)$ for $n=10$. Note that typical
    (i.e. most likely) values of $\tau_1$ are $2$ and $2n$, while the average
    value is $\langle \tau_1 \rangle = n+1$. (b) Distribution
    $P\left(\tau_1=2m\right)$ of the first revisit time to $1$.}
\label{1stRev}  
\end{figure}

The required probability is then simply the product of these two numbers
divided by the total number of walks that start at $x=1$ and are absorbed after
$2n+1$ steps, which is $C_n$. Therefore
\begin{align}
\mathbf{P}\left(2m \,| \,T\right)= \frac{C_{m-1}\,C_{n-m}}{C_n}= \frac{n+1}{m\,(n-m+1)}\frac{\binom{2m-2}{m-1}\,\binom{2n-2m}{n-m}}{\binom{2n}{n}}\,.
\end{align}
Because $\mathbf{P}\left(\tau_1=2m \,| \,T=2n+1\right)$ is symmetric under
$m \rightarrow n+1-m$, the average value of $\tau_1$, the time of the first
revisit, conditioned on $T=2n+1$, can be immediately seen to be
\begin{align}
\langle \tau_1 \rangle_n =n+1\,.
\end{align}
The above result may also be obtained by direct calculation.  Because of the
bimodal nature of the underlying probability distribution, the average value
is very different from the typical value.  The average corresponds to the
minimum of the probability distribution (Fig.~\ref{1stRev}(a)), just as in
the arcsine law for the time of the last zero of a Brownian motion.

\subsection{Time of first return to $1$ for any $n$}

From the conditional probability $\mathbf{P}\left(2m \,| \,T\right)$, we may
now compute the joint probability $\mathcal{P}\left(2m,T\right)$:
\begin{align}
\mathcal{P}\left(2m,T\right)&=\mathbf{P}\left(2m \,| \,T\right)\,P(T)\nonumber\\
&=\frac{C_{m-1}\,C_{n-m}}{C_n}\times\frac{C_n}{2^{2n}}=\frac{C_{m-1}\,C_{n-m}}{2^{2n}}.
\end{align}
With this result, we can readily obtain the distribution $P(2m)$, the
probability for a path of any length to perform an excursion of $2m$ steps
that lies above $x=2$ between steps $1$ and $2m-1$ (with the first step
constrained to go from $x=1$ to $x=2$):
\begin{align}
P(2m)&=\sum_{n\geq m}\,\mathcal{P}\left(2m, T\right)\nonumber\\
&= C_{m-1}\,\sum_{n\geq m}\,\frac{C_{n-m}}{2^{2n}}=\frac{C_{m-1}}{2^{2m-1}}\,.
\end{align}
This distribution is normalized, because $\sum_{m\geq 1} C_{m-1}/2^{2m-1}=1$.
The Markovian nature of the random walk means that there is no memory between
what happens after step $2m$ and the probability that the walk first revisits
$x=1$ at step $2m$.  Hence $P(2m)$ is simply the probability that a symmetric
random walk first returns to its starting point at step $2m$, which
asymptotically scales as $m^{-3/2}$~\cite{F64,R01,MP10}.  Because of this
scaling, the average time for the first return,
$\langle \!\langle \tau_1 \rangle\! \rangle =\sum_m 2m\,P(2m)$, is infinite,
even though $P(2m)$ is peaked at $m=1$ (Fig.~\ref{1stRev}(b)).

\section{Summary}
\label{sec:disc}

We showed how standard first-passage methods can be used to determine the
average time that a diffusing particle spends in a given spatial range when
the particle starts at some point $x_0$ and dies when it reaches an absorbing
point or set.  We also derived corresponding results for the discrete random
walk, where the analog of the residence time is the number of times that a
given point is visited.  For continuum diffusion, the average residence time
in a given spatial range is simply the integral of the probability
distribution over all time in this same range, with the given initial
condition and the absorbing boundary condition. This perspective allowed us
to also treat, in a relatively simple manner: (a) biased diffusion, (b)
diffusion in a finite interval (conditioned on absorption at a given side of
the interval), and (c) diffusion in general spatial dimensions.  It is also
worth emphasizing the time integral of the probability distribution at a
given point is essentially just the electrostatic potential at this point.
This correspondence provides a simple way to calculate residence times and to
understand the dependence of the residence time on basic parameters.

The main qualitative feature of the residence time at a given point is that
it vanishes (often linearly) in the distance between this point and the
absorber.  That is, a diffusing particle does not linger when it is close to
an absorbing point. Another interesting feature, almost intuitive from the
analogy with electrostatics, is the fact that, in low dimensions ($d\leq 3$),
the average residence time at any point beyond the starting point is constant
and simply equal to the average residence time at the starting point. This is
no longer the case for $d\geq4$, with an abrupt transition between $d=3$ and
$d=4$.  It would be of interest to understand why this transition occurs at a
spatial dimension that differs from that of the well known transition between
recurrence and transience, which happens at $d=2$.

For the discrete random walk, we exploited the generating function method to
derive parallel results for the number of times that a given lattice site is
visited before the walk dies at the absorbing point.  For a walk that starts
at $x_0=1$, there are, upon averaging over walks of all possible lengths, two
subsequent visits to $x=1$ and four subsequent visits to $x=2$ before the
random walk dies.  We also showed that the random walk makes four subsequent
visits to any point $x\geq 2$, on average.  We also found that the first
revisit to $x=1$ occurs near the start or the end of the path.  This means
that it is very unlikely that there will be a large deviation toward the
boundary and away from the average position of the path near the middle of an
excursion.  This suggests that an individual Brownian excursion always
remains close to its average shape.  We hope to investigate this behavior in
future work, with a view to obtaining a full characterization of the
fluctuations around an excursion's average semi-circular shape.

\section*{Acknowledgments}

SR wishes to acknowledge the hospitality of Universit\'e Paris-1
Panth\'eon-Sorbonne for their support of a visit during which this project
was initiated, as well as financial support from grant DMR-1608211 from the
National Science Foundation.  We also thank O.~B\'enichou and S.~N.~Majumdar
for useful discussions and advice.

\appendix

\section{Relation Between Generating Functions}

Using the geometric constraints \eref{geom} in Eq.~\eqref{genG}, the
generating function $\mathcal{G}(x,y,z)$ can be re-expressed as
\begin{align}
  \mathcal{G}(x,y,z)&=\sum_{n=1}^\infty\,\sum_{\ell=1}^n  \,\sum_{k=\ell}^n  \,
       \mathcal{P}\left(n,k,\ell\right)\,x^n\,y^k\,z^\ell \nonumber\\
                    & =\sum_{n=1}^\infty\,\sum_{\ell=1}^n \,\sum_{k=\ell}^n   \,
                      P(\ell)\,\mathbf{P}\left(n,k|\ell\right)\,x^n\,y^k\,z^\ell\,,
\end{align}
where we use $\mathbf{P}(\dots|\dots)$ to denote the conditional joint probability.

Next we write $\mathbf{P}\left(n,k|\ell\right)$ in terms of the variables
$m_i$ and $j_i$ (see Fig.~\ref{x2}):
\begin{align}
   \mathcal{G}(x,y,z)&=\sum_{n=1}^\infty\,\sum_{\ell=1}^n \,\sum_{k=\ell}^n  
  \sum_{\substack{j_1+\dots+j_\ell = k-\ell\\ m_1+\dots+m_\ell = n-\ell\\ j_i\leq m_i}}
  \hspace{-5mm} P(\ell)\,\mathbf{P}\left(m_1,\dots,m_\ell,j_1,\dots,j_\ell|\ell\right)\,x^{m_1+\dots m_{\ell}+\ell}\,y^{j_1+\dots j_{\ell}+\ell}\,z^\ell\nonumber\\
  &= \sum_{n=1}^\infty \,\sum_{\ell=1}^n\,
  \sum_{\substack{m_1+\dots+m_\ell = n-\ell\\ 0\leq j_i \leq m_i}}\,P(\ell)\,\mathbf{P}\left(m_1,\dots,m_\ell,j_1,\dots,j_\ell|\ell\right)\,x^{m_1+\dots m_{\ell}+\ell}\,y^{j_1+\dots j_{\ell}+\ell}\,z^\ell\nonumber\\
&= \sum_{\ell\geq 1}\,P(\ell)\,\sum_{n\geq \ell}\,\sum_{\substack{m_1+\dots+m_\ell = n-\ell\\ 0\leq j_i \leq m_i}}\,\mathbf{P}\left(m_1,\dots,m_\ell,j_1,\dots,j_\ell|\ell\right)\,x^{m_1+\dots m_{\ell}+\ell}\,y^{j_1+\dots j_{\ell}+\ell}\,z^\ell\nonumber\\
 & = \sum_{\ell\geq 1}\,P(\ell)\,\sum_{m_1,\dots,m_\ell\geq 0}\,
  \sum_{j_1,\dots,j_\ell=0}^{m_1,\dots,m_\ell}\,\left[\prod_{i=1}^{\ell}\,\mathbf{P}\left(m_i,j_i|\ell\right)\,x^{m_i}\,y^{k_i}\right]\,(xyz)^\ell\nonumber\\
 & = \sum_{\ell\geq 1}\,P(\ell)\,(xyz)^\ell\,\sum_{m_1,\dots,m_\ell\geq 0}\,
  \sum_{j_1,\dots,j_\ell=0}^{m_1,\dots,m_\ell}\,\left[\prod_{i=1}^{\ell}\,\mathbf{P}\left(m_i,j_i|\ell\right)\,x^{m_i}\,y^{j_i}\right]\nonumber\\
&= \sum_{\ell\geq
  1}\,P(\ell)\,(xyz)^\ell\,\prod_{i=1}^{\ell}\left[\sum_{m_i\geq
  0}\,\sum_{0\leq j_i\leq m_i}\,\,\mathbf{P}\left(m_i,j_i|\ell\right)\,x^{m_i}\,y^{j_i}\right]\,.\label{eq:fact}
\end{align}
\medskip

We now use the fact that the random walk is a Markov process, which implies
that
$\mathbf{P}\left(m_i,j_i|\ell\right)=\mathcal{P}\left(m_i,j_i\right)=A(m_i,j_i)/2^{2m_i}$.
Therefore
\begin{align}
\sum_{m_i\geq 0}\,\sum_{0\leq k_i\leq m_i}\,\,\mathbf{P}\left(m_i,j_i|\ell\right)\,x^{m_i}\,y^{j_i}=G(x,y)\,.
\end{align}
Note that $G(x,y)$, as defined in Eq.~(\ref{G}), appears here and not
$g(x,y)$ because upon starting from $x=2$, when coming from $x=1$, the path
is not conditioned to immediately move to $x=3$.  In fact, the path is allowed
to return immediately to $x=1$, as reflected
in the fact that, for the $i$th excursion, $m_i$ may be $0$.\\
Eq.~(\ref{eq:fact}) becomes
\begin{align}
\mathcal{G}(x,y,z)= \sum_{\ell\geq 1}\,P(\ell)\,(xyz)^\ell\,\prod_{1\leq
                    i\leq \ell}\left[G(x,y)\right]
=\sum_{\ell\geq 1}\,P(\ell)\,\left[x\,y\,z\,G(x,y)\right]^\ell\,.
\end{align}
This is Eq.~(\ref{calG}) in the main text.


\section*{References}
\begin{thebibliography}{99}

\bibitem{F64} W. Feller, \emph{Introduction to Probability Theory and its
    Applications, 3rd ed.}, (J Wiley \& Sons, Inc.\ New York, 1964).

\bibitem{S64} F. Spitzer, \emph{Principles of Random Walk}, (Van Nostrand, Princeton NJ, 1964)

\bibitem{MW65} E. W. Montroll and G. H. Weiss, J. Math.\ Phys.\ \textbf{6}, 167 (1965).

\bibitem{MP10} P. M\"orters and Y. Peres, \emph{Brownian Motion}, (Cambridge
  University Press, Cambridge, UK, 2010).
  
\bibitem{Chung76} K. L. Chung, Ark.\ Mat.\ \textbf{14}, 155 (1976).

\bibitem{BCC03}  A. Baldassarri, F. Colaiori, and C. Castellano,
Phys.\ Rev.\ Lett.\ \textbf{90}, 060601 (2003).

\bibitem{CBC04}  F. Colaiori, A. Baldassarri, and C. Castellano,
  Phys.\ Rev.\ E \textbf{69}, 041105 (2004).

\bibitem{L48} P. L\'evy, \emph{Processus stochastique et mouvement Brownien},
  (Gauthier-Villars, Paris, 1948).

\bibitem{R63} D. Ray, Illinois Journal of Mathematics \textbf{7}, 615 (1963)

\bibitem{K63} F. B. Knight, Trans.\ Am.\ Math,\ Soc.\ \textbf{109}, 56 (1963).

\bibitem{CTB07} S. Condamin, V. Tejedor, and O. B\'enichou, Phys.\ Rev.\ E
  \textbf{76}, 050102 (2007).

\bibitem{BD09} O. B\'enichou and J. Desbois, J. Phys.\ A: Math.\ Theor.\
  \textbf{42}, 015004 (2009).

\bibitem{BV14} O. B\'enichou and R. Voituriez, Phys.\ Repts.\ \textbf{539},
  225 (2014).

\bibitem{A84} N. Agmon, J. Chem.\ Phys.\ \textbf{81}, 3644 (1984).
  
\bibitem{D82} H. E. Daniels, Ann.\ Statist.\ \textbf{10}, 394 (1982).

\bibitem{R01} S. Redner, \emph{A Guide to First-Passage Processes},
  (Cambridge University Press, Cambridge, UK, 2001).

\bibitem{FZ01} M. Ferraro and L. Zaninetti, Phys.\ Rev.\ E \textbf{64}, 056107 (2001).

\bibitem{FZ04} M. Ferraro and L. Zaninetti, Physica A \textbf{338}, 307 (2004).

\bibitem{FZ06} M. Ferraro and L. Zaninetti, Phys.\ Rev.\ E \textbf{73}, 057102 (2006).

\bibitem{G87} J. Galambos, \emph{The Asymptotic Theory of Extreme Order
      Statistics}, (Krieger, Malabar, FL, 1987).

\bibitem{KRB10} P. L. Krapivsky, S. Redner, and E. Ben-Naim, \emph{A
      Kinetic View of Statistical Physics}, (Cambridge University Press,
    Cambridge UK, 2010).
 
\bibitem{BRB17} U. Bhat, S. Redner and O. B{\'e}nichou, J. Stat.\ Mech.\
    073213 (2017).

\bibitem{B96} D. F. Bailey, Math.\ Mag.\ \textbf{69}, 128 (1996)
  (http://www.jstor.org/stable/269067).

\bibitem{LO16} K.-H. Lee and S.-J. Oh, arXiv:1601.06685.

\bibitem{S15} R. P. Stanley, \emph{Catalan Numbers}, (Cambridge University
  Press, Cambridge UK, 2015).
  
\bibitem{S99} R. P. Stanley, \emph{Enumerative Combinatorics vol. 2},
  (Cambridge University Press, Cambridge UK, 1999).

\bibitem{IS} \emph{The On-Line Encyclopedia of Integer Sequences}, \url{https://oeis.org}.
  


\end{thebibliography}
\end{document}